\documentclass[10pt,conference]{IEEEtran}
\usepackage[noadjust]{cite}      
\usepackage{graphicx}  
\usepackage{amsmath,amssymb,epsfig}
\usepackage{multicol}
\usepackage{algorithm}
\usepackage{url}
\usepackage{float}
\usepackage{amsfonts}
\usepackage{theorem}

\makeatletter
    \newcommand\figcaption{\def\@captype{figure}\caption}
    \newcommand\tabcaption{\def\@captype{table}\caption}
\makeatother

\newtheorem{thm}{Theorem}
\newtheorem{cor}{Corollary}
\newtheorem{lem}{Lemma}

\theoremstyle{definition}

\theoremstyle{remark}

\begin{document}
\title{Performance Analysis of
Algebraic Soft Decoding of Reed-Solomon Codes over Binary Symmetric and Erasure Channels}

\author{\authorblockN{Jing~Jiang\footnotemark}
\authorblockA{Dept. of EE, Texas A\&M University,\\
College Station, TX, 77843, U.S.A\\
Email: jjiang@ee.tamu.edu} \and
\authorblockN{Krishna~R.~Narayanan}
\authorblockA{Dept of EE, Texas A\&M University,\\
College Station, TX, 77843, U.S.A\\
Email: krn@ee.tamu.edu} }
\maketitle \footnotetext[1]{This work was supported in part by Seagate Research Inc., Pittsburgh, PA, USA.}

\begin{abstract}
In this paper, we characterize the decoding region of algebraic soft decoding (ASD) \cite{koetter_kv} of
Reed-Solomon (RS) codes over erasure channels and binary symmetric channel (BSC). Optimal multiplicity
assignment strategies (MAS) are investigated and tight bounds are derived to show the ASD can significantly
outperform conventional Berlekamp Massey (BM) decoding over these channels for a wide code rate range. The
analysis technique can also be extended to other channel models, e.g., RS coded modulation over erasure
channels.
\end{abstract}


\section{Introduction}
\label{sec:introduction}

Reed-Solomon (RS) codes are powerful error correction codes, which are widely employed in many state-of-the-art
communication systems. However, in most of the existing systems, RS codes are decoded via algebraic hard
decision decoding (HDD), which does not fully exploit the error correction capability of the code.

Since the seminal works of \cite{guruswami_gs} and \cite{koetter_kv}, algebraic soft decoding (ASD) algorithms
have come to research interest due to their significant performance gain over HDD. In \cite{koetter_kv}, an
asymptotically optimal multiplicity assignment strategy (MAS) has been proposed (it is asymptotically optimal in
the sense that it maximizes the capacity \cite{koetter_kv} of ASD when the code length goes to infinity).
However, the optimal MAS and corresponding performance analysis for finite length RS codes is difficult to
obtain. Recent progress in this problem can be found in \cite{el-khamy_mas} and \cite{nayak_kv}. However, in
these papers, the proposed schemes largely rely on numerical computation and give little insight into the
decoding region of ASD. In \cite{koetter_additive}, a general framework has been studied for channels with
additive cost \cite{koetter_additive}. However, it is still interesting and of practical value to characterize
the decoding regions for particular channels.

In this paper, we first investigate the optimal MAS for ASD for
binary erasure channel (BEC) and binary symmetric channel (BSC)
and characterize the corresponding decoding region of ASD. It is
shown that ASD has a significant gain over HDD for BEC and BSC.
The analysis technique is then extended to some other discrete
alphabet channels (DAC), e.g., RS coded modulation transmitted
over erasure channels. ASD is shown to consistently outperform
conventional erasure decoding, which suggests the potential
application of ASD to practical systems, such as DSL systems as
described in \cite{wesel_robust}.

\section{Performance Analysis of ASD over BEC}
\label{sec:rs_bec}

We first consider the case when RS codewords are transmitted as bits through a BEC with erasure probability
$\epsilon$.

In the multiplicity assignment state, since the true \emph{a
posterior probability} (APP) of each bit is not available, we take
the reliability information observed from the channel, i.e, if one
bit is erased in a symbol, we regard two candidate symbols as
being equally probable. Furthermore, similar to \cite{koetter_kv}
and \cite{nayak_kv}, we assume that the codewords are uniform over
$\textrm{F}^N_q$. Consequently, there is no preference for some
symbols over others and the multiplicity assigned to equally
probable candidate symbols are assumed to be the same.

We define each symbol that has a number of $i$ bits erasures as
being of type $i$. Consequently, for a code over $GF(2^m)$, there
are $(m+1)$ types of symbols. Let the number of symbols of type
$i$ in a received codeword be $a_i$. As discussed above, we will
assign equal multiplicity to symbols of the same type; whereas,
the multiplicity assigned to type $i$ may vary according to the
received codeword. Define the multiplicity assigned to each
candidate of symbol of type $i$ as $m_i$. Thus, the total
multiplicity assigned to one symbol of type $i$ is $2^i\times
m_i$. According to the above notation, the score and cost can be
defined as in \cite{koetter_kv} as follows:

\begin{equation}
\label{eqn:score_defn} S = \sum_{i=0}^{m}{a_im_i}.
\end{equation}

\begin{equation}
\label{eqn:cost_defn} C = \sum_{i=0}^{m}{a_i\times 2^i\times {{m_i+1}\choose{2}}}\doteq
\frac{1}{2}\sum_{i=0}^{m}{a_i\times 2^i\times m_i^2}
\end{equation}

The approximation in (\ref{eqn:cost_defn}) becomes tight when $m_i$ becomes large.

In general, the decoding radius of ASD is difficult to characterize due to its soft decoding nature. However, it
is shown in \cite{koetter_kv} that the ASD is guaranteed to return the transmitted codeword when the following
inequality holds:
\begin{lem}\label{thm:guaranteed_decoding}
The sufficient condition for ASD to list the transmitted codeword is as follows:
\begin{equation}
\label{eqn:sufficient} S \ge \sqrt{2(K-1)C}
\end{equation}
\end{lem}
\begin{proof} \label{prf:sufficient}
This is proved in Corollary 5 in \cite{koetter_kv}.
\end{proof}

This sufficient condition becomes a tight approximation when $N$ is large. With a little bit abuse of notation,
we approximate the decoding region of ASD using the region guaranteed by (\ref{eqn:sufficient}). Upper bound and
lower bound are derived for this approximate decoding region, which facilitates the performance analysis in
practical systems. Further, we consider ASD with infinite cost such that we can relax the multiplicity from
integers to real numbers. It is justified by the fact that rational numbers are dense on the real axis and they
can always be scaled up to be integers with infinite cost (see also \cite{el-khamy_mas}).

We first investigate the optimal MAS for the BEC.

\begin{lem}\label{thm:proportional assignment}
The proportional multiplicity assignment strategy (PMAS) is optimal for BEC regardless of the received signal.
\end{lem}
\begin{proof} \label{prf:proportional_assignment}
Assume that the received codeword has $a_i$ symbols of type $i$. Since it can be readily shown that a larger
cost will always lead to a better performance, the MAS can be formulated as maximizing the score with a cost
constraint. With infinite cost, the problem is expressed as:

\begin{equation}
\label{eqn:asymptotic_b}
\nonumber \max_{\{m_i\}} \sum_{i=0}^{m}{a_im_i}
\end{equation}
~~~~~~~~~~~~~~~~~~~~~subject to $\frac{1}{2}\sum_{i=0}^{m}{a_i 2^i m_i^2}\le C_0$

This is a standard optimization problem with linear cost function and quadratic constraint. Using a Lagrange
multiplier, the new objective function becomes
\begin{equation}
\label{eqn:Lagrange} \emph{L} = -\sum_{i=0}^{m}{a_im_i}+\lambda
\left(\frac{1}{2}\sum_{i=0}^{m}{2^ia_im_i^2}-C_0 \right)
\end{equation}

Take the partial derivative with respect to $m_i$ and set it to zero. We have:
\begin{equation}
\label{eqn:Lagrange partial} \frac{\partial{\emph{L}}}{\partial{m_i}} = -a_i+\lambda 2^ia_im_i = 0
\end{equation}

Therefore we have $m_i = \frac{2^{-i}}{\lambda}$, i.e., $m_i \propto 2^{-i}$, which proves that PMAS is optimal.
\end{proof}
Since PMAS is optimal, we will from now on assume that PMAS is
used. Suppose for any received signal, it has type series as
$\{a_i\}$. Under PMAS, we assume that the total multiplicity for
each symbol as $M$. Consequently, the score is $S_0 =
\sum_{i=0}^{m} a_i 2^{-i} M = \eta M$ and the cost is $C_0 =
\frac{1}{2}\sum_{i=0}^{m} a_i 2^{-i} M^2 = \frac{1}{2} \eta M^2$,
where $\eta = \sum_{i=0}^{m} a_i 2^{-i}$ is a positive number. The
sufficient condition of (\ref{eqn:sufficient}) becomes:
\begin{align}
\label{align:sufficient_0}
S_0 &\ge \sqrt{2(K-1)C_0}\\
\label{align:eta_ieq}\eta &\ge K-1
\end{align}

In the following lemmas, we will first derive the worst case erasure pattern for ASD over BEC.

\begin{lem}\label{lem:monotone erasure}
If a received codeword can be decoded, it can always be decoded if some of the erasures are recovered.
\end{lem}
\begin{proof} \label{prf:monotone_erasure}
The proof is immediate by the fact that if some of the erasures are recovered, the score will increase and the
cost will decrease. Consequently, the sufficient condition (~\ref{eqn:sufficient}~) is still satisfied.
\end{proof}

\begin{lem}\label{lem:worst case erasure pattern}
Given $i$-bit erasures, the worst case erasure pattern for ASD under PMAS (in terms of the guaranteed decoding
region) is that all the bits are spread in different symbols as evenly as possible. That is: $(N - i + \lfloor
\frac{i}{N} \rfloor N)$ symbols contain $\lfloor \frac{i}{N} \rfloor$ bit errors and $(i - \lfloor \frac{i}{N}
\rfloor N)$ contain $\lceil \frac{i}{N} \rceil$ bit errors.
\end{lem}
\begin{proof} \label{prf:worst case erasure pattern}
The lemma can be readily verified by induction. Take two arbitrary symbols of type $i$ and $j$, if we average
the bit erasures between these two, we get two symbols of type $\lfloor \frac{i+j}{2} \rfloor$ and $\lceil
\frac{i+j}{2} \rceil$. The updated $\eta'$ can be expressed as:
\begin{equation}\label{eqn_eta}
\eta' = \eta + 2^{-\lfloor \frac{i+j}{2} \rfloor} + 2^{-\lceil \frac{i+j}{2}\rceil} - 2^{-i} - 2^{-j} \le \eta
\end{equation}
From (\ref{align:eta_ieq}), it is clear that when $\eta$ increases, the erasure pattern becomes better for ASD.
Thus, since $\eta \ge \eta'$, the later erasure pattern is worse. By repeating the above procedure, we can
finally get the worst erasure pattern for ASD under PMAS, i.e., the bit erasures are spread as evenly as
possible in different symbols.
\end{proof}

The asymptotic bit-level decoding radius $e$ (i.e., the worst case erasure pattern) for ASD  can be formulated
as a standard optimization problem:

$\min e = \sum_{i=0}^{m}{i e_i}$

s.t $e_i \ge 0, i = 0, 1, \cdots, m$

and $\sum_{i=0}^{m}{i e_i} = N, \eta \le K-1$

The above problem can be solved numerically using integer linear programming. However, for practical medium to
high rate RS codes, we can have a simpler form of the radius.

\begin{thm}\label{thm:worst case region}
The bit-level decoding radius for ASD can be expressed as: $e = 2(N-K+1)$ for $K \ge \frac{1}{2}N+1$ and $e =
3N-4(K-1)$ for $\frac{1}{4} N+1 \le K \le \frac{1}{2} N+1$.
\end{thm}
\begin{proof} \label{prf:worst_case_region}
According to Lemma \ref{lem:worst case erasure pattern}, the worst case erasure pattern is all erased bits are
spread evenly over different symbols. Assume the erased bits $e \le 2N$. Thus, the worst case erasure event
consists of $e_1$ symbols of type 1 and $e_2$ symbols of type 2. The problem can be formulated as:
\begin{align}
\label{align:opt}
\min~~~&e = e_1+2e_2\\
s.t.~~~&e_1 \nonumber\ge 0, e_2 \nonumber\ge 0, e_1+e_2 \nonumber\le N\\
and~~~&\eta = N-\frac{1}{2}e_1-\frac{3}{4}e_2 \nonumber\le K-1
\end{align}

After some straightforward manipulation, we can obtain the optimal value $e_1^{*}$, $e_2^{*}$ and $e^{*}$.

For $K \ge \frac{1}{2}N+1$, it is guaranteed that in the worst case erasure pattern, each symbol contains no
more than 1 bit erasure. The optimal value can be computed as:
\begin{align}
e_1^{*} &= 2(N-K+1), e_2^{*} = 0\\
\label{align:opt_e_a}e^{*} &= 2[N-(K-1)]
\end{align}

For $\frac{1}{4} N+1 \le K \le \frac{1}{2} N+1$, the optimal value can be expressed as:
\begin{align}
e_1^{*} &= 4(K-1)-N, e_2^{*} = 2[N-2(K-1)]\\
\label{align:opt_e_b}e^{*} &= 2[N-(K-1)]+N-2(K-1)
\end{align}
It is easy to verify that $e^{*} \le 2N$ in this case. Also note that $e^{*} \ge 2[N-(K-1)]$, which suggests as
the code rate goes lower, the bit-level decoding radius gets larger and larger.
\end{proof}

Theorem \ref{thm:worst case region} gives an FER upper bound on ASD performance under PMAS with infinite cost.

\begin{cor}\label{cor:lower_bound}
For codes of rate for $R \ge \frac{1}{2}+\frac{1}{N}$, the sufficient condition \ref{eqn:sufficient} cannot be
satisfied when there are more than $2(N-K+1)$ symbols in error.
\end{cor}
\begin{proof} \label{prf:lower_bound}
The corollary follows from (\ref{align:opt_e_a}) and Lemma \ref{lem:monotone erasure}. If there are more than
$2(N-K+1)$ symbols having erased bits, the most optimistic case is that these symbols are of type $1$. Besides,
due to (\ref{align:opt_e_a}), the sufficient condition is not satisfied and ASD can not guarantee to decode.
\end{proof}

\begin{figure}[h]
\begin{center}
\includegraphics[width=2.5in]{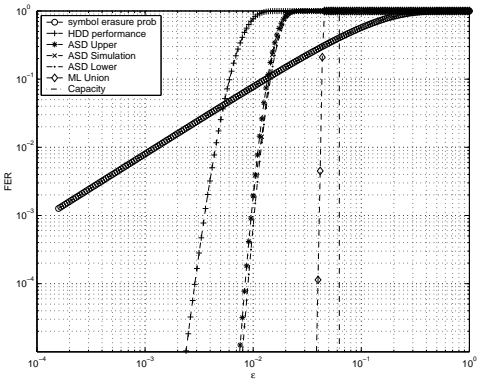}
\caption{Bounds and Simulation Results of RS(255,239) over BEC} \label{fig:upp_low_bd}
\end{center}
\end{figure}

Corollary \ref{cor:lower_bound} can serve as a lower bound of the guaranteed region of ASD. The bounds derived
in Theorem \ref{thm:worst case region} and Corollary \ref{cor:lower_bound} are shown in Figure
\ref{fig:upp_low_bd} in conjunction with the maximum likelihood (ML) performance union bound over RS averaged
ensemble \cite{retter_gs}. ML performance of RS averaged ensemble is very close to the capacity of BEC, which
suggests RS codes are good codes. Besides, ASD has a significant performance gain over conventional BM erasure
decoding. This is intuitively true, since we do not have to erase the whole symbol if part of the bits in that
symbol is erased, which can be taken advantage of by ASD. It can be seen from the figure that for practical high
rate long codes, both the upper and lower bounds are tight and they together accurately indicate the performance
of the guaranteed region of ASD down to FER$~=~10^{-20}$ (this is not shown here due to the page limit).

\section{Performance Analysis of ASD over BSC}
\label{sec:rs_bsc}

In this section, we study the performance of ASD over BSC. Since there is no bit-level soft information
available, any bit-level reliability based soft decoding algorithm does not work. However, we will show that
symbol-level soft information can still be utilized by ASD once proper MAS is applied.

Similar to the BEC case discussed in Section \ref{sec:rs_bec}, we
assume that equally probable candidate symbols are assigned equal
multiplicity. We also follow the same notations except that type
$i$ is the type that $i$ bits in that symbol get flipped. Similar
to the BEC case, the problem can be reformulated as a constrained
optimization problem:

max $S = \sum_{i=0}^{m}{a_im_i}$

$C \doteq \frac{N}{2}\sum_{i=0}^{m}{{{m}\choose{i}}\times m_i^2} \le C_{0}$,

However, unlike the BEC case, we do not know ${a_i}$ at the decoder. Consequently, the MAS should be the same
for any received codeword. Here, we resort to the asymptotically optimal MAS, i.e., we try to decode up to the
largest bit-level radius for the worst case error pattern. The problem can be viewed as a max-min problem over
$\{a_i\}$ and $\{m_i\}$.

\begin{equation}
\label{eqn:min_max} \max_{\{m_i\}}\min_{\{a_i\}} \sum_{i=0}^{m}{a_i m_i}
\end{equation}

subject to $\sum_{i=0}^{m}{a_i} = N$ where $\{a_i\}$s are integers

and $\frac{N}{2}\sum_{i=0}^{m}{{{m}\choose{i}}\times m_i^2} \le C_{0}$.

The above problem is quite complicated, since $\{a_i\}$s are integers. Even if that condition is relaxed, the
solution may only be obtained numerically, which does not give any insight into the exact decoding radius of
ASD.

We first take one step back and consider a special case of BSC, called 1-bit flipped BSC, i.e., in each symbol,
at most one bit is in error. By doing that, we only have to assign multiplicities to two types of symbols.
Suppose there are $e$ symbols in error, $e \le N$ and the score and cost are $S = (N-e)m_0 + e m_1$ and $C =
\frac{N}{2}[m_0^2 + m m_1^2]$. The asymptotically optimal MAS is just to maximize $e$. Suppose $m_1 = t m_0$,
where the multiplicity coefficient $0 \le t \le 1$, the sufficient condition in (\ref{eqn:sufficient}) becomes:

\begin{align}
\label{align:sufficient_bsc_a}[(N-e)+e t] m_0 &\ge \sqrt{(K-1)N(1+m t^2)} m_0\\
\label{align:sufficient_bsc_b}e &\le \frac{N-\sqrt{N(K-1)(1+m t^2)}}{1-t}
\end{align}

Let $t = 1$ in (\ref{align:sufficient_bsc_a}), we have the inequality independent of $e$. The inequality always
hold as long as $N \ge (K-1)(1+m)$, which is true for low rate codes, i.e., for a rate

\begin{equation}
\label{eqn:rate} R \le \frac{1}{1+m}+\frac{1}{N}
\end{equation}

In this case, setting $m_0 = m_1$ will correct all errors in the received signal. On the other hand, for higher
rate code, the optimal MAS is to optimize $t$ to maximize the RHS of (\ref{align:sufficient_bsc_b}).

The problem is equivalent to finding a point on the hyperbola $\frac{y^2}{N(K-1)}-mx^2 = 1$ within interval $x
\in [0, 1)$ such that a line passing through that point and the given point $(1, N)$ has the maximal slope. This
is nothing but the tangent to the hyperbola. For the tangential point $(x_0, y_0)$, we have the following
relationships:

\begin{equation}
\label{eqn:relation1} \frac{dy}{dx}\mid_{x = x_0} = N(K-1)m\frac{x_0}{y_0} = \frac{N-y_0}{1-x_0}
\end{equation}

\begin{equation}
\label{eqn:relation2} \frac{y_0^2}{N(K-1)}-mx_0^2 = 1
\end{equation}

We define $\Delta$ as:
\begin{equation}
\label{eqn:delta} \Delta = (m (K-1))^2+(N-K+1)(m^2(K-1)-m N)
\end{equation}

Plug in $y_0$ into (\ref{eqn:relation2}), we can get:

\begin{equation}
\label{eqn:x_0 solution} x_0 = \frac{- m (K-1)+\sqrt{\Delta}}{m^2(K-1)-m N}
\end{equation}

Using (\ref{eqn:relation1}) and (\ref{eqn:relation2}), the optimal value of the slope in terms of $x_0$ is:

\begin{equation}
\label{eqn:slope solution} d = \frac{dy}{dx}|_{x = x_0} = \frac{N m}{m+\frac{1}{x_0}}
\end{equation}

Thus $\lfloor d \rfloor$ is the exact error correction radius of ASD algorithm under asymptotically optimal MAS
over 1-bit flipped BSC. Besides, (\ref{eqn:x_0 solution}) can be further approximated for some long and high
rate codes. Observe that $(N-K+1)(m^2(K-1)-m N) \ll (m (K-1))^2$ for high rate codes, using first order Taylor
expansion, we get:

\begin{equation}
\label{eqn:taylor expansion} \sqrt{\Delta} \doteq m(K-1)[1+\frac{1}{2} \frac{(N-K+1)(m^2(K-1)-m N)}{m^2(K-1)^2}]
\end{equation}

Plug this into (\ref{eqn:x_0 solution}) and (\ref{eqn:slope solution}), we get:

\begin{equation}
\label{eqn:approximation} \tilde{d} = \frac{N(N-K+1)}{N+(K-1)}
\end{equation}

Note that the approximation becomes tight when rate is high, however, it should be noted that the performance
improvement is significant only when the rate is low. For instance, for $N = 255$, $K = 223$ ASD does not
improve the performance, for $N = 255$, $K = 167$ ASD gives an extra error correction capability over GS
decoding, for $N = 255$, $K = 77$, it corrects 7 more errors and for $N = 255$, $K = 30$, it corrects 45 more
errors. For $K < 30$, all errors can be corrected for this 1-bit flipped BSC.

Now, we show that the above MAS is also asymptotically optimal for RS codes over BSC for a wide code range.

\begin{lem}\label{thm:worst case error pattern}
Given a number of $i\le N$ bits errors and the optimal multiplicity coefficient $t \le \frac{1}{2}$, the worst
case error pattern for ASD algorithm is that all erroneous bits are spread in $i$ different symbols.
\end{lem}

\begin{proof}\label{prf:worst case error pattern}
Assume that there are $e$ bits flipped by the channel. The cost for BSC channel does not change when the MAS is
fixed. The worst case error pattern is the one that minimizes the score. In the above MAS, multiplicities are
assigned only to the received symbol and its 1-bit flipped neighbors. Thus, a potential error pattern which is
worse than the 1-bit flipped BSC will try to group bits in each symbol to reduce the score. The original score
can be expressed as:

\begin{equation}
\label{eqn:org_score} S = M[(N-e)+t e]
\end{equation}

Let the worst case error pattern has $e'$ symbols containing 1 bit error and $e''$ symbols containing more than
1-bit error. Evidently, for symbols containing more than 2-bit errors, we can always further decrease the score
by splitting these bit errors into one symbol containing 2-bit errors and the other containing the rest of the
errors. Consequently, the worst case error pattern will contain symbols with at most 2-bit errors. We have
$e'+2e'' = e$. The score becomes:

\begin{equation}
\label{eqn:2bit_score} S'' = M[(N-e'-e'') + t e'] = M[(N-e) + e t + e'' (1-2t)]
\end{equation}
When $t \le \frac{1}{2}$, $S'' \ge S$, which proves that spreading all the bits in different symbols is the
worst case error pattern.
\end{proof}

\begin{thm}\label{thm:bsc_bound}
For BSC, the asymptotic optimal MAS of ASD algorithm can guarantee to decode up to $d$ bits, where $d$ is
computed in (\ref{eqn:slope solution}), given $d \le N$ and the optimal multiplicity coefficient $t \le
\frac{1}{2}$.
\end{thm}
\begin{proof} \label{prf:bsc_bound}
According to lemma \ref{thm:worst case error pattern}, all bits spread in different symbols are the worst case
error pattern for the ASD algorithm, which is nothing but the 1-bit flipped BSC. Thus, they will be the
asymptotically dominating error patterns. On the other hand, it is shown before that the proposed MAS is
asymptotically optimal for 1-bit flipped BSC, i.e., maximizing the asymptotic decoding radius $d$. Consequently,
the proposed MAS will guarantee to decode all error patterns with fewer than $d$-bit errors over BSC as well.
\end{proof}

\begin{figure}[h]
\begin{center}
\includegraphics[width=2.5in]{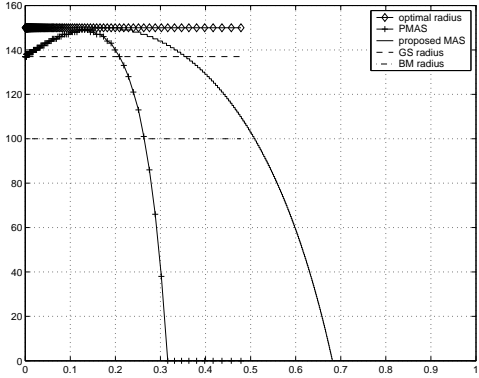}
\caption{Bit level decoding radius of an RS (255,55) code} \label{fig:radius}
\end{center}
\end{figure}

The error correction radii as a function of $t$ are given in Figure \ref{fig:radius}. It can be seen that the
optimal MAS (which is achieved by $t =0.2$) corrects 13 and 50 more bit errors than GS and BM asymptotically.
Besides, we also plot bit-level radius of PMAS, where the x-axis is the crossover probability $p_c$ of the BSC.
Note that PMAS is not asymptotically optimal for BSC. Even though we choose $p_c$ to maximize the bit-level
radius (around $p_c = 0.13$), the bit-level decoding radius is still 1 bit smaller than that of the optimal MAS.
The reason can be explained as follows: the worst case error pattern of BSC is shown to be all bit-level errors
spread in different symbols, thus, the asymptotically optimal MAS only has to assign multiplicities to symbols
of type $0$ and type $1$. On the other hand, KV algorithm assigns multiplicities proportionally. Thus it also
assigns multiplicities to candidate symbols with more than 1-bit flip, which makes it suboptimal in terms of
bit-level decoding radius.


We consider the performance of this asymptotically optimal MAS in Figure \ref{fig:rs(15,3)}. Consider AWGN
channel with 1-bit quantization before decoding, which is equivalent to BSC. The ASD under the proposed MAS
outperforms GS decoding and HDD by 1.6dB and 3.5dB respectively at an FER = $10^{-5}$.

\begin{figure}[h]
\begin{center}
\includegraphics[width=2.5in]{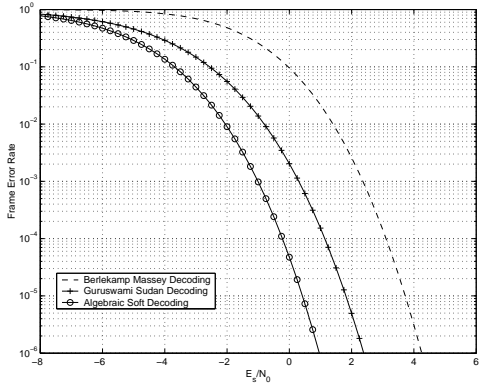}
\caption{Performance Comparison for RS (15,3) code} \label{fig:rs(15,3)}
\end{center}
\end{figure}


\section{Performance Analysis of ASD of RS Coded Modulation over Erasure Channels}
\label{sec:rs_modulation}

We extend the result from Section \ref{sec:rs_bec} to RS coded modulation transmitted over erasure channels,
i.e. $u$ bits of coded information are grouped together and transmitted using a $2^u$-ary QAM or PSK modulation
format. The channel will erase the signal with erasure probability $\epsilon$. Practical channels of this model
are discussed in \cite{wesel_robust}.

In Section \ref{sec:rs_bec}, we showed that PMAS is optimal for erasure channels. Clearly, all erasure patterns
in this coded modulation model is a subset of the erasure patterns of BEC. Thus, PMAS is also optimal for this
channel model. Let each PSK (or QAM) symbol contains $u$ bits. Without loss of generality, we assume $u$ divides
$m$, i.e., $m = lu$. Thus, for each symbol, we have $(l+1)$ types.

\begin{lem}\label{thm:worst case modulation}
The worst case erasure pattern for ASD under PMAS is that all erasure events are spread in different symbols as
evenly as possible.
\end{lem}

\begin{proof} \label{prf:worst case modulation}
Assume two RS symbols are of type $i$ and $j$, we can average all the erasures events between the two symbols,
we have:
\begin{align}
\label{align:eta_rs_mod} \eta' = \eta +2^{-\lfloor \frac{i+j}{2} \rfloor u}+2^{-\lceil \frac{i+j}{2} \rceil
u}-2^{-iu}-2^{-ju} \le \eta
\end{align}
Similar to Lemma \ref{lem:worst case erasure pattern}, when $\eta$ decreases, spreading erasure events in
different RS symbols evenly is the worst case.
\end{proof}

\begin{thm}\label{thm:bounds_memory}
ASD under PMAS can guarantee to decode up to $(N-K+1)/(1-2^{-u})$ erasure events if $R \ge 2^{-u}+\frac{1}{N}$.
\end{thm}
\begin{proof} \label{prf:bounds_memory}
According to Lemma \ref{thm:worst case modulation}, spreading erasure events in different symbols is the worst
case erasure pattern if $K \ge 2^{-u}N+1$.

Assume that $e$ symbols of type $1$ are erased and the rest $(N-e)$ symbols are of type $0$. Thus $\eta =
N-(1-2^{-u})e$. According to (\ref{align:eta_ieq}) when the following condition is satisfied:
\begin{align}
\label{align:region} e \le (N-K+1)/(1-2^{-u})
\end{align}
ASD is guaranteed to decode the transmitted codeword.
\end{proof}
Note that this asymptotic decoding radius is consistently larger than the conventional RS erasure decoding
region, $e = N-K$ erasure events. Note that (\ref{align:region}) is a generalization of (\ref{align:opt_e_a}) in
Theorem \ref{thm:worst case region} (with $u = 1$ as special case).

\section{Conclusion}
\label{sec:conclusion}

We have presented optimal multiplicity assignment strategies and performance analysis of algebraic soft decision
decoding over erasure channels and BSC. It was shown that ASD under optimal MAS can significantly outperform the
BM and GS algorithm.


\section*{Acknowledgment}
The authors are grateful to N.~Ratnakar and R.~Koetter for many insightful discussions. They also thank
anonymous reviewers for their constructive comments that greatly improve the quality of this paper.



%


\end{document}